\documentclass[aps, prl, 10pt, twocolumn,preprintnumbers,amsmath,amssymb]{revtex4}
\usepackage{amssymb}
\usepackage{amsmath}
\usepackage{graphicx}
\usepackage{color}

\begin{document}
\title{Atomically thin semiconductors as nonlinear mirrors}

\author{Sina Zeytinoglu}
\affiliation{Institute for Quantum Electronics, ETH Z\"urich, CH-8093 Zurich, Switzerland.}
\author{Charlaine Roth}
\affiliation{Institute for Quantum Electronics, ETH Z\"urich, CH-8093 Zurich, Switzerland.}
\author{Sebastian Huber}
\affiliation{Institute for Quantum Electronics, ETH Z\"urich,
CH-8093 Zurich, Switzerland.}
\author{Atac \.Imamo\u{g}lu}
\affiliation{Institute for Quantum Electronics, ETH Z\"urich, CH-8093 Zurich, Switzerland.}

\date{\today}

\begin{abstract}
We show that a transition metal dichalcogenide monolayer with a
radiatively broadened exciton resonance would exhibit perfect
extinction of a transmitted field. This result holds for $s$- or $p$-polarized weak resonant light fields at any incidence angle, due to the conservation
of in-plane momentum of excitons and photons in a flat defect-free
two dimensional crystal. In contrast to extinction experiments with
single quantum emitters, exciton-exciton interactions lead to an
enhancement of reflection with increasing power for incident fields
that are blue detuned with respect to the exciton resonance.
We show that the interactions limit the maximum reflection that can be
achieved by depleting the incoming coherent state into an outgoing
two-mode squeezed state.
\end{abstract}

% 03.67.Lx: Quantum computation
% 03.67.Hk: Quantum communication
% 73.21.La: Electron states and collective excitations in Quantum Dots
% 42.50.-p: Quantum optics
% 78.67.Hc:
% 42.50.Ex:
% 78.30.Fs:
\maketitle

Monolayers of transition metal dichalcogenides (TMD) such as MoSe$_2$ or WSe$_2$ constitute a new class of two dimensional (2D) direct band-gap semiconductors~\cite{ Radisavljevic2011, Splendiani2010, Baugher2014, Britnell2013}. Lowest energy elementary optical excitations in TMDs in the absence of free electrons or holes are excitons with an ultra-large binding energy of $\sim 0.5$~eV~\cite{Chernikov2014}. Remarkably, recent experiments have demonstrated predominantly spontaneous-emission limited exciton transition linewidths in clean MoSe$_2$ flakes embedded in hexagonal boron nitride (hBN) layers~\cite{Back2017}. Since radiative broadening dominates over disorder induced inhomogeneous broadening, TMD monolayers can be considered as ideal two-dimensional (2D) optical materials.

In this Letter we show that a TMD monolayer acts as a perfect atomically-thin mirror for radiation that is resonant with radiatively broadened excitonic resonances. In the limit of weak resonant incident laser field, destructive interference between the transmitted field and the field generated by the TMD excitons leads to perfect extinction. In-plane momentum conservation ensures that the transmitted field vanishes for any incidence angle as long as the generated 2D excitons have a perfect overlap with the incident field polarization: this is the case for an incident $s$-polarized field. Remarkably, p-polarized fields also yield perfect extinction since the generated longitudinally-polarized exciton couples exclusively to $p$-polarized outgoing radiation. On the other hand, any superposition of $s$- and $p$-polarized fields will have finite transmission: this is a consequence of finite energy splitting of the transverse and longitudinal exciton resonances induced by the electron-hole exchange interaction \cite{Yu2014}.

The exciton-exciton interactions ensure that there will be a non-zero transmitted field as the intensity of the field is increased: this is the analog of saturation induced reduction in extinction observed in resonantly driven single quantum emitters. Unlike the latter however, strong extinction of the mean transmitted field is possible in the TMD case by tuning the incident field to blue (red) side of the resonance for repulsive (attractive) exciton-exciton interactions. We find that for laser detunings and intensities where the exciton system approaches bistability, the extinction of the mean field is only limited by the depletion of the mean excitonic field due to interactions. The transmitted field in this regime is a broadband squeezed coherent state.

%%%%%%%%%%%%%%%%%%%%%%
\begin{figure}[t!]
\includegraphics{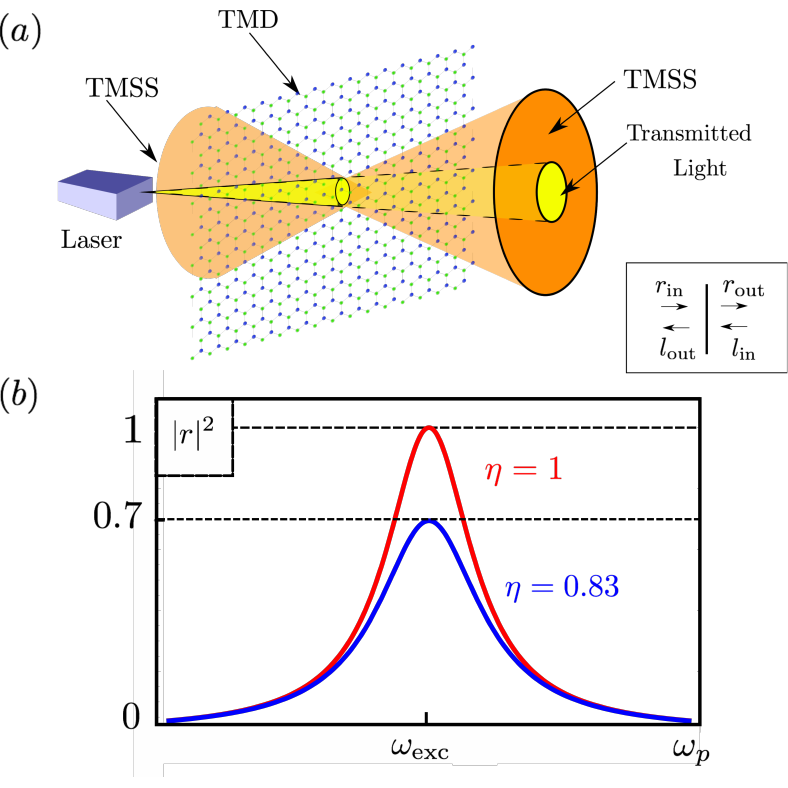}
\caption{(a) The schematic of the experimental setup. A collimated coherent laser field is incident on a transition metal dichalcogenide (TMD) monolayer. In addition to coherent transmitted and reflected fields, a two-mode squeezed-state (TMSS) is generated due to exciton-exciton interactions. Unlike the coherent fields, the TMSS is emitted into a large solid-angle. Electromagnetic field can be characterized as consisting of right (left) propagating input $r_{\mathrm{in}}$ ($l_{\mathrm{in}}$) and output $r_{\mathrm{out}}$ ($l_{\mathrm{out}}$) modes. (b) The reflection from the 2D TMD layer
for weak near resonant drive. Destructive interference between the directly transmitted field and the field generated by the TMD excitons lead to a suppression of transmission and an enhancement of reflection. For perfect coupling efficiency $\eta=1$ between the input light and the excitons on the TMD layer, the system constitutes a perfect mirror. Imperfect coupling efficiency due to
the non-radiative lifetime of the excitons reduces the reflection maxima to $\eta^2$.
}
\label{fig:Setup}
\end{figure}

Figure~$\ref{fig:Setup}$ (a) depicts the experimental system we analyze. We assume near-resonant monochromatic light incident on a monolayer TMD, whose excitonic excitations have the ladder operators $x_k$. We model the electromagnetic environment that the excitonic field couples to in terms of right and left propagating field modes whose ladder operators are denoted as $r_{k}$ and $l_k$, respectively. We assume a defect-free flat TMD monolayer and a large excitation spot such that in-plane momentum $k$ is conserved, and the total Hamiltonian can be written as
\begin{align}
H  = \sum H(k) =
\sum_k \left[ H_{\mathrm{TMD}}(k) + H_{\mathrm{bath}}(k) + H_{\mathrm{int}}(k)\right],
\end{align}
where
\begin{align}
H_{\mathrm{bath}}(k) &=
\int_{-\infty}^{\infty} d\omega \omega
\left[ r_k^{\dagger}(\omega) r_k(\omega) + l_k(\omega) l_k(\omega)\right],
\\
H_{\mathrm{int}}(k) &=
\int_{-\infty}^{\infty} d\omega \kappa  e^{i\theta/2} x_k^{\dagger}
\left[ r_k(\omega)+l_k(\omega) +h.c. \right],
\end{align}
and $H_{\mathrm{TMD}}$ is to be specified. To simplify the expressions, we have set $\hbar = 1$ and expressed frequencies in a frame rotating with the incident coherent field frequency  $\omega_p$. Parameters $\kappa$ and $\theta/2$ are the strength and the phase of the coupling, respectively. The bath operators are properly normalized using the appropriate density of states [e.g., $r_{k}(\omega) =\sqrt{\rho_{\mathrm{env}}(k,\omega)}r_{k}(k_z)$ ]. The density of states of the environment depends on the in-plane momentum and frequency in the rotating frame of $\omega_p$ as $\rho_{\mathrm{env}}(k,\omega)=\frac{1}{c}\frac{\omega+\omega_p}{\sqrt{(\omega+\omega_p)^2 - (c k) ^2}}$. Similarly the coupling constant $\kappa(\omega)$ also includes a factor $\sqrt{\rho_{\mathrm{env}}(k,\omega)}$. In the following, we only consider in-plane wave-vectors for which the electromagnetic bath can be treated as Markovian.
%We note that the density of states for the bath degrees of freedom depends strongly on frequency in the close
%vicinity of the excitonic wavenumber $k = (\omega_0+\omega)/c$.
%
%$x_k(\omega) =\sqrt{ \frac{m}{\hbar^2}}x_k(k_z)$ and the bath degrees of freedom $r_{k}(\omega) =\sqrt{\frac{\omega+\omega_0}{\sqrt{(\omega+\omega_0)^2 - (c k) ^2}}}r_{k}(k_z)$

The Heisenberg-Langevin equation of motion for $x_k$ evolving under the influence of such a bath is given by
\begin{align}
\nonumber \dot{x}_k(t) &= \frac{i}{\hbar}[H_{\mathrm{TMD}},x_k(t)] \\
&- \left\{ i e^{-i\theta}\sqrt{\gamma}\left[r_k^{\mathrm{in}}(t) + l_k^{\mathrm{in}}(t)\right] + \frac{2 \gamma}{2} x_k(t) \right\},
\label{eq:HeisLang}
\end{align}
where $\gamma  \equiv 2\pi \kappa^2$ is the radiative decay rate of TMD excitons into the right (left) moving modes in the Markov approximation. In Eq.~($\ref{eq:HeisLang}$), the right (left) moving input noise operator, $r_k^{\mathrm{in}}$ $(l_k^{\mathrm{in}})$ is defined as
\begin{align}
r_k^{\mathrm{in}}(t) = \int_{-\infty}^{\infty} e^{-i\omega(t-t_0)} r_k(\omega) d\omega,
\label{eq:inputop}
\end{align}
with $t_0<t$. The input-output relation in frequency space is \cite{walls2008quantum}
\begin{align}
r_k^{\mathrm{out}}(\omega) = r_k^{\mathrm{in}}(\omega) - i e^{i \theta} \sqrt{\gamma}x_k(\omega),
\label{eq:IandO}
\end{align}
where $\omega$ is the frequency label of the bath modes, and $r_{\mathrm{out}}(\omega)$ is defined the same way as in Eq.~($\ref{eq:inputop}$) but with $t_0 \leftrightarrow t_f>t$. The input-output relation allows us to express the output fields in terms of the input fields once we solve the equation of motion of $x_k$ in terms of the input noise operators \cite{walls2008quantum}

We first consider a non-interacting exciton system where $H_{\mathrm{TMD}}(k) = \omega_{\mathrm{exc}}(k) x_{k}^{\dagger}x_{k}$, with $\omega_{exc}(k)=\omega_{exc} + k^2/(2m_{exc})- \omega_p$, where $\omega_{\mathrm{exc}}$,
is the exciton energy in the lab frame.
We then obtain the following solution to Eq. ($\ref{eq:HeisLang}$)
\begin{align}
x_{k}(\omega) = -i \left(e^{-i\theta}\sqrt{\gamma} \right)G_0(\omega,k )\left[r_k^{\mathrm{in}}(\omega) + l_k^{\mathrm{in}}(\omega)\right],
\label{eq:ExNonInt}
\end{align}
where we introduced the free propagator $G_0(\omega,k )~\equiv~\frac{1}{\gamma- i[(\omega - \omega_\mathrm{exc}(k)]}$.
Using Eq.~($\ref{eq:IandO}$), the solution for the right outgoing field can be written as
\begin{align}
r_k^{\mathrm{out}}(\omega) = \left[1- \gamma G_0(\omega , k)\right] r_k^{\mathrm{in}}(\omega)- \gamma G_0(\omega , k)l_k^{\mathrm{in}}(\omega).
\label{eq:SoSimp}
\end{align}
The quantity $\gamma G_0(\omega , k)$ can also be understood
as the reflection coefficient of the TMD system. Remarkably, at resonance [$\omega=\omega_{exc}(k)$], the propagator is purely real and equal to the density of states $G_0(\omega_k , k)~=~\gamma^{-1}$. Hence, the expectation value of $r_{k}^{\mathrm{out}}(\omega)$ over a right moving coherent field is zero and the reflection is perfect, cf. Fig.~\ref{fig:Setup} (b). For each $k \ll (\omega+\omega_p)/c$, $H_{\mathrm{int}}(k)$ describes a single harmonic oscillator mode coupled to a one-dimensional radiation field reservoir with $\rho_{\mathrm{env}}(k,\omega) \simeq 1/c$. The transmisison/reflection problem is therefore formally equivalent to the corresponding problem for a single quantum emitter coupled to an optical or microwave waveguide \cite{Wallraff2004}, in the limit of a weak incident coherent field where anharmonicity can be neglected.

The effect of coupling to additional decay channels can be taken into account by $\gamma\rightarrow \bar{\gamma}$.
In this case, we obtain at resonance
\begin{align}
\nonumber r_k^{\mathrm{out}}(\omega_k) &\approx  \left[1- \frac{\gamma}{\bar{\gamma}}\right]r_{k}^{\mathrm{in}}(\omega_k) - \frac{\gamma}{\bar{\gamma}}l_{k}^{\mathrm{in}}(\omega_k)  \\
\equiv& (1-\eta) r_{k}^{\mathrm{in}}(\omega_k) - \eta l_{k}^{\mathrm{in}}(\omega_k),
\end{align}
where $\eta \equiv \gamma/\bar{\gamma}$ is the coupling efficiency of the TMD excitons to the electromagnetic modes,
with the identical in-plane momentum, $k$.
As shown in Fig. $\ref{fig:Setup}$~(b),  the peak reflectivity at resonance is given by $\eta^2$.

When the input coherent field is incident on the TMD layer at an angle $\theta_{\mathrm{inc}}\neq 0 $ the
polarization of the incoming light becomes important. In particular, the polarization
component perpendicular to the TMD layer does not couple to $x_{k}(t)$.
As a result, the coupling constant for the $p$ polarized light is reduced with respect to
the $s$ polarized light by a factor  $\cos^2(\theta_{\mathrm{inc}})$. 
The role of polarization can be described,
by using an additional polarization label $\epsilon = s , p$ describing polarization of the incoming light.
Then the input output relation is written as
\begin{align}
r_{k,\epsilon}^{\mathrm{out}}(\omega) \equiv \left[1- \gamma_{\epsilon} G_0(\omega , k)\right] r_{k,\epsilon}^{\mathrm{in}}(\omega),
\end{align}
where again we assumed that the left moving input modes are in the
vacuum state. Moreover, electron-hole exchange interactions in a TMD
monolayer ensure that the excitonic excitations that couple to $s$
and $p$ polarized light are spectrally distinct and lead to
parabolic (linear) dispersion for $s$ ($p$) polarized excitons
\cite{Yu2014}. Even though both $s$ and $p$ polarized
incident light would exhibit perfect extinction on their respective
resonances, the reflectivity is below unity for any $\omega_p$ when
the incoming light is in a superposition of the $s$ and $p$
polarizations.

The effects of exciton-exciton interactions can be taken into account by
\begin{align}
\nonumber H^{\mathrm{int}}_{\mathrm{TMD}} &= \left[ \sum_{k}  \omega_{\mathrm{exc}}(k) x_{k}^{\dagger}x_{k} \right] + \frac{g}{2} x^{\dagger}_{0}x^{\dagger}_{0}x_{0}x_{0} \\
&+ \frac{g}{2}\sum'_{k} (  x^{\dagger}_{k}x^{\dagger}_{-k}x_{0}x_{0} + x^{\dagger}_{0}x^{\dagger}_{0}x_{k}x_{-k}+4 x^{\dagger}_{0}x_{0}x^{\dagger}_{k}x_{k}),
\end{align}
where we kept the highest order terms that would dominate the dynamics in the limit of a large coherent amplitude in the driven exciton mode $k=0$. We assumed repulsive contact interactions $g>0$ and excluded the $k=0$ mode in the sum $\sum_{k}'$. We first obtain the density of particles at $k=0$ when the system is driven by a coherent plane wave that is normally incident on the TMD plane [$\langle r_{k=0}^{in} (\omega=0) \rangle = \beta \sqrt{\gamma}$]: solving the Heisenberg-Langevin equation and substituting in the mean field at $k=0$, we obtain
\begin{align}
\langle x_{0}\rangle = -i \left( e^{-i\theta} \gamma \beta \right)\bar{G}(0,0)
\label{eq:intex}
\end{align}
where $\bar{G}(\omega,k)=\frac{1}{\gamma- i \left[\omega + \bar{\omega}(k)\right]} $ with $\bar{\omega}_k =  \omega_{\mathrm{exc}}(k)+(2-\delta_{k,0})g |\psi_0|^2 $, and
$|\psi_0|^2 \equiv \langle x_{0}^{\dagger}x_{0}\rangle$ is the number of excitons at mode $k=\omega=0$.
The excitonic density obeys the following self-consistency equation
\begin{align}
|\psi_0|^2 =\gamma^2 |\bar{G}_0|^2 |\beta|^2 = \frac{\gamma^2}{\gamma^2 + \bar{\omega}_{0}^2} |\beta|^2,
\label{eq:DensCohSta}
\end{align}
which is  third order in $|\psi_0|^2$ and results in
bistable solutions for $(\omega - \omega_{0})  > \sqrt{3}\gamma$ \cite{Ciuti2005} .

\begin{figure}[ht]
\centering
\includegraphics[width=0.8\columnwidth]{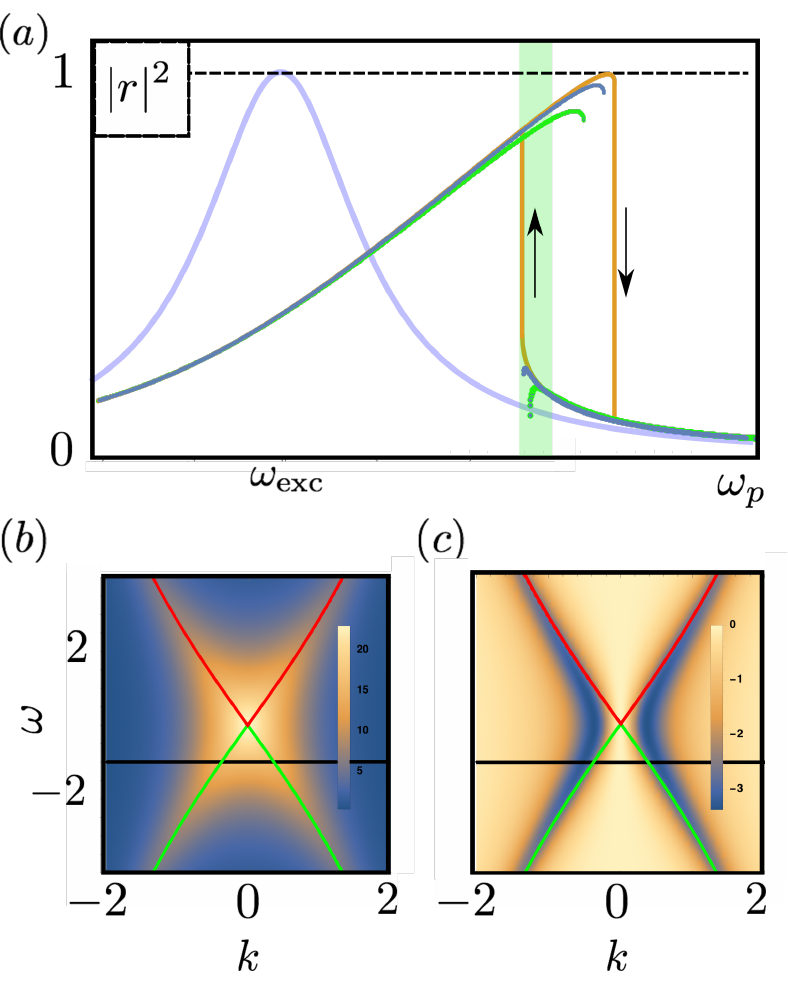}
\caption{(a) The reduction of reflection due to depletion of the $k=0$ coherent
state for three different values of $g$ while keeping $g\beta^2=3.5$ constant.
All parameters are plotted in the units of $\gamma$.
In the limit that $g\rightarrow 0 $ (orange line), the depletion does not modify the reflection maximum,
since in this limit the excitonic density at resonance is infinite.
For $g=1.75$ ($g=0.35$) [green (blue) line], the depletion reduces the reflection maximum down
to $\approx 0.9$ ($\approx 0.97$).
The green shaded area indicates the region where
the mean field treatment is expected to be not reliable for this choice of $g$. Here, a
depletion of the coherent exciton density exceeding 10 $\%$ is used as a benchmark.
The black arrows indicate the hysteresis curve. The faint blue curve is the same as Fig. $\ref{fig:Setup} (b)$ for $\eta=1$. The anti-squeezing (b) and squeezing (c) spectra at $\Delta= g\beta^2 = 3.5$,
in units of $\gamma$. For this detuning and coherent exciton density, the fluctuations on top of the coherent field have a Bogoliubov-like spectrum of an interacting bosonic system. The bandwidth of squeezing is determined by the decay rate of the excitons. While there is no upper bound for anti-squeezing in the
$g\rightarrow 0$ limit, the largest squeezing that can be achieved is limited to $\approx3$ dB
since the excitonic system is coupled to two separate baths.}
\label{fig:Squeeze}
\end{figure}
In the mean-field limit where the second line of Eq.~(11) is neglected, the condition for perfect extinction is $\bar{\omega}_0~=~0$. As expected, perfect extinction entails that the exciton flux through the TMD layer equals the photon influx. That is,
\begin{align}
\gamma |\tilde{\psi}_0|^2= \gamma |\beta|^2 =  \gamma \frac{\omega_p - \omega_{\mathrm{exc}}(k=0)}{g} ,
\label{eq:perfext}
\end{align}
where $|\tilde{\psi}_0|^2$ is the density of excitons at perfect extinction.
We emphasize that in this mean-field limit, perfect extinction is possible for \textit{any} positive (negative) value of $g$ in the TMD monolayer as long as the pump is blue (red) detuned by an amount determined by Eq.~($\ref{eq:perfext}$) from the bare excitonic resonance. Naturally, the expression in Eq. ($\ref{eq:perfext}$) interpolates to the non-interacting case when both $g$ and the detuning is taken to zero. Figure~$\ref{fig:Squeeze}$ (a) shows the mean-field reflection spectrum (yellow) as $\omega_p$ is tuned across the resonance for $g \beta^2 = 3.5$: the excitonic system is bistable for this choice of parameters and the unity reflection can be obtained as $\omega_p$ is tuned across the bare excitonic resonance towards higher frequencies.

%\textcolor{red}{The equation of motion for the excitonic mode $x_k$ then has the frequency space solution
%\begin{align}
%\nonumber  x_{k}(\omega) =& -i  G_0(\omega,k)\times\bigg\{ \left(e^{-i\theta} \sqrt{\gamma}\right) \left[ r_k^{\mathrm{in}}(\omega) + l_k^{\mathrm{in}}(\omega)\right] \\
%&+ g x_0^2 x_{-k}^{\dagger}(-\omega)+ 2 g  x_0^{\dagger}x_0 x_{k}(\omega)\bigg \},
%\label{eq:exc}
%\end{align}}
%To obtain the semiclassical solution, we displace the
%input operator $r_{k}^{\mathrm{in}}(\omega=0) = \beta \sqrt{\gamma} + \delta r_{k}^{\mathrm{in}}(\omega=0)$,
%where the fluctuations $\delta r$ vanish when averaged over the coherent state.
%Keeping terms only in the first order of fluctuations Eq.
% ($\ref{eq:exc}$) describes the evolution of the fluctuations on top of the excitonic coherent
% state in the reference frame of the the pump laser. We note that in this notation $\omega_k \neq \omega_{-k}$
% for all $k_p \neq 0 $, and from now on, we pick $k_p = 0$ for simplicity.

To take into account the effect of quantum fluctuations on top
of the coherent excitonic field $\psi_0$, we solve the equations of motion for these fluctuations up to linear order.
In our analysis, we assume a single parabolic dispersion ($s$-polarized) excitonic dispersion for simplicity.
The solution for the $k~\neq~0$ excitonic operators are
  \begin{align}
 \nonumber x_{k}(\omega)  &= -i\left(e^{-i\theta}\sqrt{\gamma}\right)G(\omega, k)\\
& \times \left\{n_k^{\mathrm{in}}(\omega) +  U[\bar{G}(-\omega,-k)]^* (n_{-k}^{\mathrm{in}})^{\dagger}(-\omega)\right \},
 \label{eq:ExDressed}
 \end{align}
 where $n_{k}^{\mathrm{in}}(\omega) \equiv r_k^{\mathrm{in}}(\omega) + l_k^{\mathrm{in}}(\omega)$, $U \equiv  - ig \psi_0^2$, and the dressed propagator is $$G(\omega,k) \equiv \frac{[\bar{G}^{-1}(-\omega, -k)]^{*}}{\bar{G}^{-1}(\omega,k)  [\bar{G}^{-1}(-\omega, -k)]^*-|U|^2}. $$
 Eq.~(\ref{eq:ExDressed}) allows for a clear interpretation when contrasted with the solution of the non-interacting problem in Eq.~($\ref{eq:ExNonInt}$). The first difference between the two is the extra noise term $ U [\bar{G}(-\omega,-k)]^* (n_{-k}^{\mathrm{in}})^{\dagger}$, which describes
 the particle number violating coupling between the excitons and the fluctuations of the electromagnetic vacuum.
 Microscopically, this is allowed due to the interactions mediated by the excitonic coherent state.
 Secondly, both noise terms are now propagated by the dressed propagator $G(\omega,k)$. We note that the imaginary
 part of the poles of $G(\omega,k)$ gives the excitation spectrum for the dissipative
 excitonic condensate, while the real part of the poles gives the dissipation rate of the corresponding modes
 [see Fig. $\ref{fig:Squeeze}$ (b-c)]. We note that the functions $G(k,\omega)$ and $G(k,\omega) U
 [\bar{G}(-\omega,-k)]^*$ can be thought of as the normal and anomalous propagators for the
weakly interacting excitonic condensate coupled to a Markovian bath.

Using the input-output relation, we find the solution for the right moving output operator
\begin{align}
\nonumber r_k^{\mathrm{out}}(\omega) &= u_{k}(\omega) r_{k}^{\mathrm{in}}(\omega) + v_{k}(\omega) (r_{-k}^{\mathrm{in}})^{\dagger}(-\omega)\\
 &+ \bar{u}_{k}(\omega) l_{k}^{\mathrm{in}}(\omega) +\bar{v}_{k}(\omega)  (l_{-k}^{\mathrm{in}})^{\dagger}(-\omega),
\label{eq:fullso2}
\end{align}
where
\begin{align}
u_{k}(\omega) &= 1- \bar{u}_{k}(\omega) = 1-\gamma G(\omega_k)\\
v_{k}(\omega) &= \bar{v}_{k}(\omega) = \gamma G(\omega_k) U [\bar{G}(-\omega,-k)]^* .
\label{eq:cohfac}
\end{align}
%\begin{align}
%\nonumber r_k^{\mathrm{out}}(\omega) &= \left[1-\gamma G(\omega,k)\right]r_{k}^{\mathrm{in}}(\omega)  - \gamma G(\omega,k) l_{k}^{\mathrm{in}}(\omega)\\
%&-\gamma \frac{V}{\bar{G}^{-1}_0(\omega, k )\left[\bar{G}_0^{-1}(-\omega,-k)\right]^{*} - |V|^2}(n_{-k}^{\mathrm{in}})^{\dagger}(-\omega),
%\label{eq:fullsol}
%\end{align}
The form of the solution in Eq. ($\ref{eq:fullso2}$) ensures that $\langle (r_k^{\mathrm{out}})^{\dagger}(\omega) r_k^{\mathrm{out}}(\omega)\rangle \neq 0$ even for $k\neq 0$. Physically, this means that
the input coherent population at $k = 0$ is depleted out into output modes with $k\neq 0$.
As a result, the perfect reflection condition in Eq. ($\ref{eq:perfext}$) cannot be satisfied
even at resonance [i.e., $\bar{\omega}(0) = 0$].
Thus, the reduction of reflection can be viewed as a renormalization of the
\textit{decay rate} $\gamma$ of the coherent density of excitons with $k~=~0$.

The conservation of input and output photon fluxes in the absence of absorption yields
\begin{align}
\sum_{\sigma = r, l}\iint d \omega d^2 k\langle (\sigma^{\mathrm{out}})^\dagger(\omega) \sigma^{\mathrm{out}}(\omega)\rangle-\langle (\sigma^{\mathrm{in}})^\dagger(\omega) \sigma^{\mathrm{in}}(\omega)\rangle = 0.
\end{align}
Evaluating the expectation values, and defining the reflection coefficient $r$
self-consistently gives us the set of equations sufficient to solve for $r$ and $\bar{\gamma}$
\begin{align}
|r|^2-{\rm Re}(r)+D &= 0 \label{eq:OptTh}\\
r -\frac{\gamma}{\bar{\gamma}- i[\omega_{\mathrm{exc}}(0) + g|r|^2 \beta^2]}&=0
\end{align}
where
\begin{align}
2D\equiv \frac{1}{2\pi^3} \iint _{-\infty}^{\infty}d^2k d\omega \frac{|v_{k}(\omega) |^2}{ \gamma \beta^2}
\label{eq:depletion}
\end{align}
is the total quantum depletion into the right and left propagating output modes. The parameter
$\bar{\gamma} = \bar{\gamma}_{D}>\gamma$ is the renormalized radiative decay rate of the coherent density
of excitons at $k=0$, and ensures
that the optical theorem in Eq.~(\ref{eq:OptTh}) is satisfied.

%We note that thermal depletion does not result in further reduction of extinction.
%Assuming the depletion is small, $\eta \approx 4 \iint _{-\infty}^{\infty}d k d\omega \frac{|v_{k}(\omega) |^2 }{\gamma \beta^2}$.
%An intuitive way to estimate the effect is by realizing that
%in the steady state the photon flux through the TMD layer needs to be constant.
%The depletion associated with the squeezing that is given by
%\begin{align}
%\iint _{-\infty}^{\infty}dk d\omega \left\langle \left[r_k^{\mathrm{out}}(\omega)\right]^{\dagger}r_k^{\mathrm{out}}(\omega)\right \rangle = 2 \iint _{-\infty}^{\infty}d k d\omega |v_{k}(\omega) |^2,
%\end{align}
%and similarly for the $l_{k}^{\mathrm{out}}(\omega)$ modes. At any moment, the depletion of the particles
% from the condensate result in a reduction of the excitonic coherent state. This, in turn, results in the
% reduction of the extinction according to the relation
% \begin{align}
%\nonumber \iint_{-\infty}^{\infty}dk d\omega\langle(r_k^{\mathrm{in}})^\dagger r_k^{\mathrm{in}}\rangle + \langle(l_k^{\mathrm{in}})^\dagger l_k^{\mathrm{in}}\rangle= \\
% \iint_{-\infty}^{\infty}dk d\omega \langle(r_k^{\mathrm{out}})^\dagger r_k^{\mathrm{out}}\rangle + \langle(l_k{^\mathrm{out}})^\dagger l_k^{\mathrm{out}}\rangle,
% \end{align}
% \textcolor{red}{since the output density of the left moving modes is given by $\gamma\langle x_0^{\dagger}x_0 \rangle$}

The quantum depletion is significant in the vicinity of the excitonic bistability  threshold \cite{Ciuti2005},
where the real part of the pole of the dressed propagator [$G(k,\omega)$]
approaches zero, causing $v_k(0)$ defined in Eq. ($\ref{eq:cohfac}$) to diverge.
However, in the limit where $g$ is taken to zero while keeping $g\beta^2$ constant, the
region where the depletion is significant shrinks down to a point at the bistability threshold.
This effect can also be understood as a consequence of the $\gamma\beta^2$ term in the
denominator of Eq.~($\ref{eq:depletion}$). The reduction of reflectivity for two different values of
$g$ with appropriately scaled pump power is shown in Fig.~$\ref{fig:Squeeze}$~(a).
In the figure, we also indicate the region of the pump frequency $\omega_p$ where the depletion
of the excitonic condensate is a significant fraction of the excitonic density (green shaded region); in this parameter range,
the mean field treatment is expected to be unreliable.
%approaches zero from below, increasing the lifetime of the fluctuations on top of the excitonic
%coherent state [see Fig. $\ref{fig:Squeeze}$ (d-e)]. This is accompanied by an increase in the two-mode squeezing
%and the depletion of particles from the coherent state at $k=0$.

Eq. (\ref{eq:fullso2}) also describes the squeezing of the right moving output noise. The correlations between the $\pm k$ modes manifest in
the fact that the fluctuations in the output field are superpositions of input noise with $\pm k$.
Using the expressions above, we can calculate the variance in the quadrature fluctuations
\begin{align}
\nonumber (\Delta X^2_{\pm})(k,\omega) &\equiv \frac{1}{2}\langle || (r_{-k}^{\mathrm{out}})^{\dagger}(-\omega)\pm r_k^{\mathrm{out}}(\omega)||^2\rangle,
\end{align}
where $||o||^2 \equiv o^{\dagger}o$. We plot the squeezing and anti-squeezing spectra at perfect extinction in
Fig.~$\ref{fig:Squeeze}$~(b~-~c), where we consider the limit $g \rightarrow 0$ while keeping $g\beta^2$ constant.
We observe that at resonance (i.e, $\bar{\omega}(0) = 0$), the dispersion of the fluctuations follow the equilibrium
Bogoliubov dispersion. Because the excitons are coupled to both right and left propagating baths, the squeezing
properties of the output field is similar to those of a non-degenerate optical parametric oscillator (OPO) implemented
by a two sided cavity \cite{gardiner2004quantum,Reynaud1988}.

Our results suggest that a TMD monolayer could constitute an interesting source of anti-squeezed light with favourable properties. Even though small exciton Bohr radius implies that the interaction strength is small compared to that of GaAs excitons, the four-wave-mixing process leading to parametric down-conversion is triply resonant implying that the conversion efficiency could be sizeable. Here, translational invariance can be used to filter out the pump field. Since the nonlinear medium is atomically thin, there are no phase- or mode-matching conditions. Most importantly, the squeezing bandwidth, determined by $\gamma$, is larger than $250$~GHz, which is orders of magnitude larger than the bandwidth of the squeezed vacuum generated in cavity based optical parametric oscillators.

The Authors acknowledge many useful discussions with I. Carusotto, C. Ciuti, M. Wouters and M. Van Regemortel on squeezed light generation in driven nonlinear exciton and polariton systems. This work is supported by an ERC Advanced investigator grant
(POLTDES) and the Swiss National Science Foundation.

\bibliographystyle{unsrt}

\end{document}